\documentclass[manuscript,screen]{acmart}

\AtBeginDocument{%
  \providecommand\BibTeX{{%
    \normalfont B\kern-0.5em{\scshape i\kern-0.25em b}\kern-0.8em\TeX}}}

\setcopyright{acmcopyright}
\copyrightyear{2022}
\acmYear{2022}
\acmDOI{XXXXXXX.XXXXXXX}

\usepackage{graphicx}
\usepackage{multirow}
\usepackage{times}
\usepackage{amsmath,amsthm}
\usepackage{algorithm}
\usepackage{mathrsfs}
\usepackage{booktabs} 
\usepackage{hyperref} 
\usepackage{url}
\usepackage{algpseudocode}
\usepackage{pifont}
\usepackage{bm}
\usepackage{comment}
\usepackage{wasysym}
\usepackage{bbding-zy}
\usepackage[figuresright]{rotating}

\def\1{{\bf{1}}}
\def\0{{\bf{0}}}

\def\a{{\bf a}}
\def\b{{\bf b}}

\def\g{{\bf g}}
\def\h{{\bf h}}

\def\p{{\bf p}}
\def\q{{\bf q}}

\def\s{{\bf s}}

\def\u{{\bf u}}

\def\x{{\bf x}}

\def\z{{\bf z}}

\def\W{{\bf W}}

\def\Ecal{{\mathcal{E}}}

\def\Lcal{{\mathcal{L}}}

\def\Ncal{{\mathcal{N}}}

\def\Tcal{{\mathcal{T}}}

\def\Vcal{{\mathcal{V}}}

\def\Rbb{{\mathbb R}}

\def\todo{\textcolor{blue}}

\begin{document}

\title{Stock Movement Prediction Based on Bi-typed Hybrid-relational Market Knowledge Graph via Dual Attention Networks}



\author{Yu Zhao}
\affiliation{%
  \streetaddress{Fintech Innovation Center, Financial Intelligence and Financial Engineering Key Laboratory of Sichuan Province, Department of Artificial Intelligence}
  \institution{Southwestern University of Finance and Economics}
  \city{Chengdu}
  \country{China}
}
\email{email: zhaoyu@swufe.edu.cn}


\author{Huaming Du}
\affiliation{%
  \streetaddress{School of Business Administration, Faculty of Business Administration}
  \institution{Southwestern University of Finance and Economics}
  \city{Chengdu}
  \country{China}}

\author{Ying Liu}
\affiliation{%
  \institution{Southwestern University of Finance and Economics}
  \city{Chengdu}
  \country{China}}

\author{Shaopeng Wei}
\affiliation{%
  \streetaddress{School of Business Administration, Faculty of Business Administration}
  \institution{Southwestern University of Finance and Economics}
  \city{Chengdu}
  \country{China}}

\author{Xingyan Chen}
\affiliation{%
  \institution{Southwestern University of Finance and Economics}
  \city{Chengdu}
  \country{China}}


\author{Fuzhen Zhuang}
\affiliation{%
  \streetaddress{Institute of Artificial Intelligence}
  \institution{Beihang University}
  \city{Beijing}
  \country{China}
  }
\email{email: zhuangfuzhen@buaa.edu.cn}


\author{Qing Li}
\affiliation{%
  \streetaddress{Fintech Innovation Center, Financial Intelligence and Financial Engineering Key Laboratory}
  \institution{Southwestern University of Finance and Economics}
  \city{Chengdu}
  \country{China}
  }
\email{email: liq_t@swufe.edu.cn}

\author{Ji Liu}
\affiliation{%
  \institution{AI Lab, Kwai Inc.}
  \city{Seattle}
  \state{Washington}
  \country{USA}
  \postcode{(He is named one MIT technology review’s “35 innovators under 35 in China”, Google Scholar Citation\ >\ 8000))}
  }
\email{email: ji.liu.uwisc@gmail.com}

\author{Gang Kou}
\authornote{Gang Kou is the corresponding author.}
\email{email: kougang@swufe.edu.cn}
\affiliation{%
  \streetaddress{School of Business Administration, Faculty of Business Administration}
  \institution{Southwestern University of Finance and Economics}
  \city{Chengdu}
  \country{China}
  \postcode{(Professor Kou awards the National Science Fund for Distinguished Young Scholars, Google Scholar Citation\ >\ 11000).}
  }
\renewcommand{\shortauthors}{Y. Zhao, H. Du, Y. Liu et al.}
\renewcommand{\shorttitle}{Stock Movement Prediction Based on Bi-typed Hybrid-relational MKG via Dual Attention Networks.}

\begin{abstract}
Stock Movement Prediction (SMP) aims at predicting listed companies' stock future price trend, which is a challenging task due to the volatile nature of financial markets. 
Recent financial studies show that the momentum spillover effect plays a significant role in stock fluctuation. However, previous studies typically only learn the simple connection information among related companies, which inevitably fail to model complex relations of listed companies in real financial market.
To address this issue, we first construct a more comprehensive Market Knowledge Graph (MKG) which contains bi-typed entities including listed companies and their associated executives, and hybrid-relations including the explicit relations and implicit relations.
Afterward, we propose \textsc{DanSmp}, a novel Dual Attention Networks to learn the momentum spillover signals based upon the constructed MKG for stock prediction.
The empirical experiments on our constructed datasets against nine SOTA baselines demonstrate that the proposed \textsc{DanSmp} is capable of improving stock prediction with the constructed MKG. 
\end{abstract}



\maketitle

\section{Introduction}
\label{sec:introduction}

%
%
%
%

 

Stock Movement Prediction (SMP) 
is a hot topic in \textit{Fintech} area since investors continuously
attempt to predict the stock future trend of listed companies for seeking maximized profit in the volatile financial market \cite{Li2017Web,Hu2018Listening,Cheng2021Modeling,Wang2021Coupling}. 
The task has spurred the interest of researchers over the years to develop better predictive models \cite{Wang2021Coupling}. 
In particular, the application of machine learning approaches yields a promising performance for SMP task \cite{Feng2019Temporal,Ye2020Multi-Graph}. 
Previous studies in both finance and AI research fields predicting a stock movement
rely on time-series analysis techniques 
using its own historical prices 
(e.g. \textit{opening price, closing price, volume}, etc)
\cite{Lin2017Hybrid,Feng2019Enhancing}. 
According to the Efficient Market Hypothesis (ETH) that implies financial market is informationally efficient \cite{Malkiel1970Efficient}, therefore, besides these stock trading factors, other researchers 
mine more indicative features from its outside-market data such as web media \cite{Li2017Web}, including news information \cite{Ming2014Stock,Liu2018Hierarchical,Li2020A} and social media \cite{Bollen2011Twitter,Si2013Exploiting,Nguyen2015Sentiment}, while ignoring the stock fluctuation diffusion influence from its related companies, which is also known as momentum spillover effect \cite{Ali2020Shared} in finance. 


Recent studies attempt to model stock momentum spillover via Graph Neural Networks (GNN) \cite{Velickovic2018Graph}. However, most of them only consider the simple explicit relations among related companies \cite{Feng2019Temporal,Ye2020Multi-Graph,Sawhney2020Spatiotemporal,Li2020Modeling}, which inevitably fail to model the complex connections of listed companies in real financial market, such as the implicit relation, and the associated executives-based meta-relations \cite{cai2016price,jing2021online}.

\begin{figure}[t]
    \centering
    \includegraphics[width=0.8\textwidth]{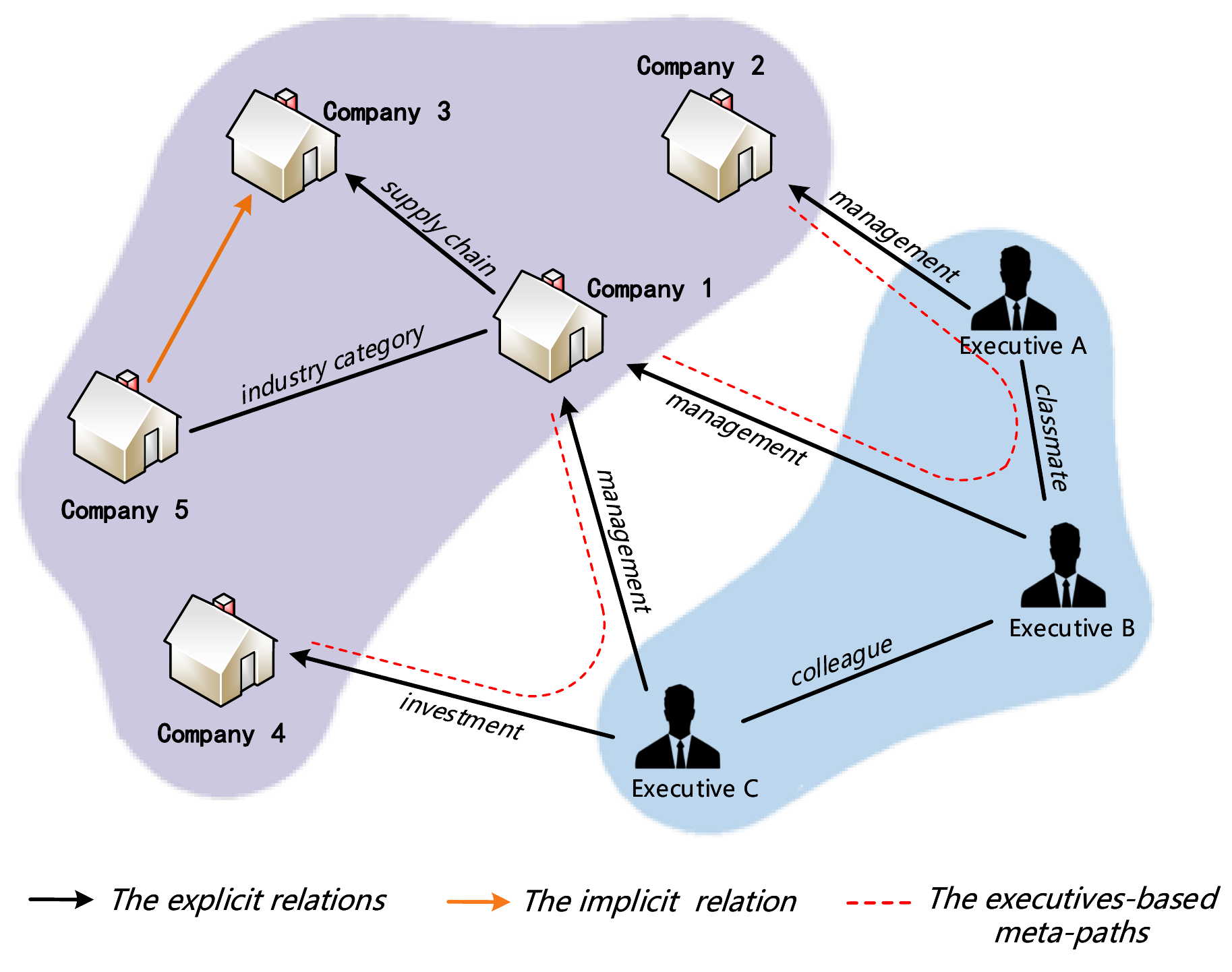}
    \caption{Example of a bi-typed (i.e. \textit{listed companies}, and \textit{executives}) hybrid-relational (i.e. \textit{explicit relations}, and \textit{implicit relation}) market knowledge graph (MKG).
    The relational information in MKG is essential for stock prediction but has not been well utilized in previous works.
    }
    \label{figure-instance}
\end{figure}

To address this issue, we construct a more comprehensive {M}arket {K}nowledge {G}raph (MKG), which consists of a considerable amount of triples in the form of ({\emph{head entity}}, {\emph{relation}}, {\emph{tail entity}}),
indicating that there exists a relation between the two entities. 
Different from previous graphs in other SOTA works \cite{Chen2018Incorporating,Feng2019Temporal,Ye2020Multi-Graph,Li2020Modeling,Sawhney2021Stock,Cheng2021Modeling}, the newly constructed MKG develops two essential characteristics: \textbf{(1)} \textit{\textbf{Bi-typed}}, i.e. containing the significant associated executive entities aside from the ordinary company entities; \textbf{(2)} \textit{\textbf{Hybrid-relational}}, i.e. providing an additional implicit relation among listed companies aside from their typical explicit relations. 
Figure $\ref{figure-instance}$ shows a toy example of MKG (See Section \ref{section-mkgc} for more details).


Afterward, to learn the stock\footnote{The term ``listed company" and ``stock" are used interchangeably.} momentum spillover signals on such bi-typed hybrid-relational MKG for stock movement prediction, we pertinently propose a novel \textbf{D}ual \textbf{A}ttention \textbf{N}etworks (\textsc{DanSmp}), as shown in Figure \ref{figure-DanSmp-Model}-II.
Specifically, the proposed model \textsc{DanSmp} is equipped with dual attention modules that are able to learn the inter-class interaction among listed companies and associated executives, and their own complex intra-class interaction alternately. 
Different from previous methods that can only model homogeneous stock graph \cite{Chen2018Incorporating,Cheng2021Modeling} or heterogeneity of stock explicit relations \cite{Nelson2017Stock,Chen2019Investment,Sawhney2021Stock}, our method is able to learn bi-typed heterogeneous entities and hybrid-relations in newly constructed market graph of stock for its spillover effects. 
The comprehensive comparison between the existing state-of-the-art (SOTA) methods with our newly proposed \textsc{DanSmp} model in terms of used market signals and main ideas is shown in Table \ref{table-model-comparison-characteristics}, demonstrating the distinguished advantage of our work.

We collect public data and construct two new SMP datasets (named \textbf{CSI100E}\footnote{"E" denotes an extension version.} and \textbf{CSI300E}) based on Chinese Stock Index to evaluate the proposed method, since no existing benchmark datasets can satisfy our need. 
Aside from the typical stock historical prices and media news, our newly published benchmark datasets CSI100E and CSI300E also provide rich market knowledge graph as mentioned above.
The empirical experimental results on CSI100E and CSI300E against nine SOTA methods demonstrate the better performance of our model \textsc{DanSmp} with MKG. 
The ablation studies reaffirm that the performance gain mainly comes from the use of the associated executives, and additional implicit relation among companies in MKG via the proposed \textsc{DanSmp}.

The {contributions} of this paper are threefold:

\begin{itemize}
    \item To model stock momentum spillover via the complex relations among companies in real market, we first construct a novel market knowledge graph. To the best of our knowledge, this study is the first attempt to explore such bi-typed hybrid-relational knowledge graph of stock via heterogeneous GNNs for its spillover effects.
    
    \item We then propose \textsc{DanSmp}, a novel Dual Attention Networks to learn the stock momentum spillover features based on the newly constructed bi-typed hybrid-relational MKG for stock prediction, which is also a non-trivial and challenging task.
    
    \item We propose two new benchmark datasets (CSI100E and CSI300E) to evaluate our method, which are also expected to promote Fintech research field further. The empirical experiments on our constructed datasets demonstrate our method can successfully improve stock prediction with bi-typed hybrid-relational MKG via the proposed \textsc{DanSmp}\footnote{The source code and our newly constructed benchmark datasets (CSI100E and CSI300E) will be released on \url{Github}: {https://github.com/trytodoit227/\textsc{DANSMP}}}.
\end{itemize}


The rest of the paper is organized as follows. In Section \ref{section-related-work}, we summarize and compare the related work. In Section \ref{section-market-signal}, we introduce the market signals for stocks prediction. Section \ref{section-method} introduces the details of the proposed methodology. 
Extensive experiments are conducted to evaluate the effectiveness of the proposed model in Section \ref{section-experiments}. Finally, we conclude the paper in Section \ref{section-conclusion}.

\section{Related Work}
\label{section-related-work}

In this section, we evaluate the existing relevant research on stock prediction. 
Stock movement prediction (SMP) has received a great deal of attention from both investors and researchers since it helps investors to make good investment decisions \cite{Rather2015Recurrent,Ding2016Knowledge-driven,Li2016A,Deng2019Knowledge-Driven}. In general, traditional SMP methods mainly can be categorized into two classes: {technical analysis} and {fundamental analysis}, according to the different types of the available stock own information they mainly used. Another major aspect for yielding better stock prediction is to utilize the stock connection information \cite{Chen2018Incorporating,Ye2020Multi-Graph,Sawhney2020Spatiotemporal,Cheng2021Modeling}. We review them in the following.

\subsection{Technical Analysis}
Technical analysis takes time-series historical market data of a stock, such as trading price and volume, as features to make prediction \cite{Edwards2018Technical,Chen2019Investment}. The basic idea behind this type of approach is to discover the hidden trading patterns that can be used for SMP. Most recent methods of this type predict stock movement trend using deep learning models \cite{Nelson2017Stock,Bao2017A,Lin2017Hybrid}. 
To further capture the long-term dependency in time series, the Recurrent Neural Networks (RNN) especially Long Short-Term Memory networks (LSTM) have been usually leveraged for prediction \cite{Gao2016Stock}. 
\citet{Bao2017A} presented a deep learning framework for stock forecasting using stacked auto-encoders and LSTM. 
\citet{Nelson2017Stock} studied the usage of LSTM networks to predict future trends of stock prices based on the price history, alongside with technical analysis indicators. 
\citet{Lin2017Hybrid} proposed an end-to-end hybrid neural networks that leverage convolutional neural networks (CNNs) and LSTM to learn local and global contextual features respectively for predicting the trend of time series. 
\citet{Zhang2017Stock} proposed a state frequency memory recurrent network to capture the multi-frequency trading patterns for stock price prediction. 
\citet{Feng2019Enhancing} proposed to employ adversarial training and add perturbations to simulate the stochasticity of price variable, and train the model to work well under
small yet intentional perturbations. 
Despite their achieved progress, however, technical analysis faces an issue that it is incapable of unveiling the rules that govern the fluctuation of the market beyond stock price data.

\subsection{Fundamental Analysis}
On the contrary, fundamental analysis takes advantage of information from outside market price data, such as economic and financial environment, and other qualitative and quantitative factors \cite{Hu2018Listening,Zhang2018Improving,Xu2018Stock}. Many methods are proposed to explore the relation between stock market and web media, e.g., news information, and social media opinion \cite{Li2017Web,Akita2016Deep,Zhang2018Improving,Wu2018Hybrid}. For instance, 
\citet{Ming2014Stock} mined text information from Wall Street Journal for SMP. 
\citet{Akita2016Deep} presented a deep learning method for stock prediction using numerical and textual information. 
\citet{Vargas2017Deep} proposed a deep learning method for stock market prediction from financial news articles. 
\citet{Xu2018Stock} put forward a novel deep generative model jointly exploiting text and price signals for this task. 
\citet{Liu2018Hierarchical} presented a hierarchical complementary attention network for predicting stock price movements with news. 
\citet{Li2020A} proposed a multimodal event-driven LSTM model for stock prediction using online news. 
Some researchers mined market media news via analyzing its sentiment, and used it for SMP \cite{Sedinkina2019Automatic,Qin2019What}. For instance, \citet{Bollen2011Twitter} analyzed twitter mood to predict the stock market. 
\citet{Si2013Exploiting} exploited topics based on twitter sentiment for stock prediction. 
\citet{Nguyen2015Sentiment} incorporated the sentiments of the specific topics of the company into the stock prediction model using social media. 
\citet{Rekabsaz2017Volatility} investigated the sentiment of annual disclosures of companies in stock markets to forecast volatility. 
\citet{Qin2019What} proposed a multimodal method that takes CEO’s vocal features, such as emotions and voice tones, into consideration. 
In this paper, we extract the signals from stock historical prices and media news sentiments as the sequential embeddings of stocks.

\subsection{Stock Relations Modeling}
\label{Section-srm}
Recent SMP studies take stock relations into consideration \cite{Chen2018Incorporating,Sawhney2020Spatiotemporal,Li2020Modeling,Cheng2021Modeling}.
For instance, 
\citet{Chen2018Incorporating} proposed to incorporate corporation relationship via graph convolutional neural networks for stock price prediction. 
\citet{Feng2019Temporal} captured the stock relations in a time-sensitive manner for stock
prediction. 
\citet{Kim2019HATS} proposed a hierarchical attention network for stock prediction using relational data. 
\citet{Li2019Multi-task} presented a multi-task recurrent neural network (RNN) with high-order Markov random fields (MRFs) to predict stock price movement direction using stock’s historical records together with its correlated stocks. 
\citet{Li2020Modeling} proposed a LSTM Relational Graph Convolutional Network (LSTM-RGCN) model, which models the connection among stocks with their correlation matrix. 
\citet{Ye2020Multi-Graph} encoded multiple relationships among stocks into graphs based on financial domain knowledge and utilized GCN to extract the cross effect based on these pre-defined graphs for stock prediction. 
\citet{Sawhney2020Spatiotemporal} proposed a spatio-temporal hypergraph convolution network for stock movement forecasting. 
\citet{Cheng2021Modeling} proposed to model the momentum spillover effect for stock prediction via attribute-driven graph attention networks. 
Despite the substantial efforts of these SOTA methods, surprisingly, most of them only focus on modeling the momentum spillover via the explicit relations among stocks, while ignoring their complex relations in real market.


Table \ref{table-model-comparison-characteristics} summarizes the key advantages of our model, comparing with a variety of previous state-of-the-art (SOTA) studies in terms of the used market signals, their methods and GNN types. (1) Different from previous studies, our method takes advantage of all three types of stock market signals, including stock historical data, media news, and market knowledge graph. In particular, we construct a more comprehensive heterogeneous market graph that contains explicit relations, implicit relations and executive relations. (2) Different from most existing models that can only {model homogeneous stock graph} \cite{Chen2018Incorporating,Cheng2021Modeling}, or heterogeneity of stock explicit relations \cite{Feng2019Temporal,Sawhney2020Deep,Ye2020Multi-Graph,Sawhney2021Stock}, which fall down in modeling heterogeneity of entities in real market, we propose a novel dual attention networks that is able to model bi-typed heterogeneous entities and hybrid-relations in newly constructed market graph of stock for its spillover effects. (3) To the best of our knowledge, this work is the first attempt to study stock movement prediction via heterogeneous GNNs.

\begin{sidewaystable}[thp]
~\\ ~\\ ~\\ ~\\ ~\\ ~\\ ~\\ ~\\ ~\\ ~\\ ~\\ ~\\ ~\\ ~\\ ~\\ ~\\ ~\\ ~\\ ~\\ ~\\ ~\\ ~\\ ~\\ ~\\ ~\\ ~\\ ~\\ ~\\ ~\\ ~\\ ~\\ ~\\ ~\\ ~\\ ~\\ ~\\ ~\\~\\ 
    \begin{center}
         \caption{Comparison between several SOTA methods and the proposed model in terms of used market signals and main ideas.}
         \label{table-model-comparison-characteristics}
         \newcommand{\tabincell}[2]{\begin{tabular}{@{}#1@{}}#2\end{tabular}}
        \resizebox{\textwidth}{!}{
          \begin{tabular}{l|l|c|c|c|c|c|c|c|c|c}
            \toprule
            \multirow{3}*{\bf{Literature} } & \multirow{3}*{\bf{Main ideas} }& \multirow{3}*{\bf{Market} } & \multirow{3}*{\bf{Metrics} } & \multirow{3}*{\bf{Method} }&
            \multirow{3}*{\bf{GNN Types}}&
            \multicolumn{5}{c}{\textbf{Stock Market Signals}}    \\
            \cline{7-11}
             &  & & & & &\multirow{2}*{\tabincell{c}{\textbf{Historical}\\ \textbf{Data}} }&\multirow{2}*{\tabincell{c}{\textbf{Media}\\ \textbf{News}} } & \multicolumn{3}{c}{\textbf{Market Knowledge Graph}} \\
             \cline{9-11}
             &  & & & & & & &\tabincell{c}{\textbf{Explicit}\\ \textbf{Relation}} & \tabincell{c}{\textbf{Implicit}\\ \textbf{Relation}} &\tabincell{c}{\textbf{Executives}\\ \textbf{Relation}}\\
            \midrule
            \midrule
            \tabincell{l}{\textbf{EB-CNN}\\ \cite{ding2015deep}}&\tabincell{l}{$\bullet$ Neural tensor network for \\ \quad learning event embedding \\ $\bullet$ Deep CNN to model the \\ \quad combined influence} &\textbf{S\&P500} & \tabincell{c}{\textbf{DA},\\ \textbf{MCC},\\ \textbf{Profit}}&\tabincell{c}{\textbf{Open IE},\\ \textbf{CNN}} & \textbf{-}&\XSolidBrush & \Checkmark & \XSolidBrush & \XSolidBrush &\XSolidBrush \\     
            \midrule
            \tabincell{l}{\textbf{SFM}\\ \cite{Zhang2017Stock}} & \tabincell{l}{$\bullet$ Extending LSTM by decomposing \\ \quad the hidden memory states \\ $\bullet$ Modeling the latent trading \\ \quad patterns with multiple frequencies } &\textbf{NASDAQ} & \textbf{Average square error}& \tabincell{c}{\textbf{LSTM},\\ \textbf{DFT}} &\textbf{-}&\Checkmark & \XSolidBrush & \XSolidBrush & \XSolidBrush &\XSolidBrush \\
            \midrule
            \tabincell{l}{\textbf{Chen's}\\ \cite{Chen2018Incorporating}} & \tabincell{l} {$\bullet$ Single investment relation} &\textbf{CSI300} & \textbf{DA}&\tabincell{c}{\textbf{LSTM},\\ \textbf{GCN}} &\textbf{Homogeneous GNNs}& \Checkmark & \XSolidBrush & \Checkmark & \XSolidBrush &\XSolidBrush \\
            \midrule
            \tabincell{l}{\textbf{TGC}\\ \cite{Feng2019Temporal}} &\tabincell{l}{$\bullet$ Single historical data \\ $\bullet$ Dynamically adjust predefined\\ \quad firm relations } &\tabincell{c}{\textbf{NASDAQ},\\ \textbf{NYSE}} & \tabincell{c}{\textbf{MSE},\\ \textbf{MRR},\\ \textbf{IRR}}& \tabincell{c}{\textbf{LSTM},\\ \textbf{Temporal Graph}\\ \textbf{Convolution}}&\textbf{Homogeneous GNNs}& \Checkmark & \XSolidBrush & \Checkmark &\XSolidBrush & \XSolidBrush \\
            \midrule
            \tabincell{l}{\textbf{HATS}\\ \cite{Kim2019HATS}}& \tabincell{l}{$\bullet$ Hierarchical aggregate different \\ \quad types of firm relational data } &\textbf{S\&P500} & \tabincell{c}{\textbf{SR},\\ \textbf{F1},\\ \textbf{DA},\\ \textbf{Return}}& \textbf{GAT}&\textbf{Homogeneous GNNs}& \Checkmark & \XSolidBrush & \Checkmark & \XSolidBrush &\XSolidBrush\\
            \midrule

             \tabincell{l}{\textbf{ALBERT+eventHAN}\\ \cite{wu-2020-event}}&\tabincell{l}{$\bullet$ ALBERT enhanced event\\ \quad  representations \\ $\bullet$ Event-enhanced hierarchical\\ \quad  attention network }&\tabincell{c}{\textbf{S\&P500},\\ \textbf{DOW},\\ \textbf{NASDAQ}} & \tabincell{c}{\textbf{DA},\\ \textbf{Annualized return}}&\tabincell{c}{\textbf{Open IE},\\ \textbf{ALBERT},\\ \textbf{HAN}} & \textbf{
            -}& \XSolidBrush & \Checkmark & \XSolidBrush & \XSolidBrush &\XSolidBrush \\     
            \midrule
            \tabincell{l}{\textbf{MAN-SF}\\ \cite{Sawhney2020Deep}}  & \tabincell{l}{$\bullet$ Multi-modal market information \\ $\bullet$ Hierarchical graph attention method } & \textbf{S\&P500} & \tabincell{c}{\textbf{F1},\\ \textbf{MCC}}& \tabincell{c}{\textbf{GAT},\\ \textbf{GRU}}& \textbf{Homogeneous GNNs}& \Checkmark & \Checkmark & \Checkmark & \XSolidBrush &\XSolidBrush\\
            \midrule
            \tabincell{l}{\textbf{STHAN-SR}\\ \cite{Sawhney2021Stock}} & \tabincell{l}{$\bullet$ A neural hypergraph architecture \\ \quad for stock selection \\ $\bullet$ Temporal hawkes attention \\ \quad mechanism } & \tabincell{c}{\textbf{NASDAQ},\\ \textbf{NYSE},\\ \textbf{TSE}} & \tabincell{c}{\textbf{SR},\\ \textbf{IRR},\\ \textbf{NDCG\@5}} & \tabincell{c}{\textbf{Hypergraph} \\ \textbf{Convolution}} & \textbf{Homogeneous GNNs}& \Checkmark & \Checkmark & \Checkmark & \XSolidBrush &\XSolidBrush\\
            
            \midrule
            \tabincell{l}{\textbf{AD-GAT}\\ \cite{Cheng2021Modeling}} & \tabincell{l}{$\bullet$ Multi-modal market information \\ $\bullet$ Attribute-driven graph \\ \quad attention network }&\textbf{S\&P500} & \tabincell{c}{\textbf{DA},\\ \textbf{AUC}}&  \textbf{GAT}&\textbf{Homogeneous GNNs}&  \Checkmark & \Checkmark & \XSolidBrush & \Checkmark &\XSolidBrush\\
            \midrule
            
            \textbf{\textsc{DanSmp} (ours)} &\tabincell{l}{$\bullet$ Multi-modal market information \\ $\bullet$ Bi-typed hybrid-relation data \\ $\bullet$ Dual attention network} &\tabincell{c}{\textbf{CSI100E},\\ \textbf{CSI300E}} & \tabincell{c}{\textbf{DA},\\ \textbf{AUC},\\ \textbf{SR},\\ \textbf{IRR}}& \tabincell{c}{\textbf{GRU},\\ \textbf{Dual Attention Network}} &\textbf{Heterogeneous GNNs}& \Checkmark & \Checkmark & \Checkmark & \Checkmark  &\Checkmark \\
            
            \bottomrule
          \end{tabular}
         }
    \end{center}
\end{sidewaystable}

\section{Market Signals}
\label{section-market-signal}
In this section, we introduce the significant signals of stocks in real financial market. We first give the details of the newly proposed bi-typed hybrid-relational market knowledge graph. Afterward, we introduce the  historical price data and media news of stocks. Most of previous works focus on partial financial market information, which makes their modeling insufficient. In this work, we take advantage of all three types of market data, fusing numerical, textual, and relational data together, for stock prediction. The features of each data used in this study are summarized in Table \ref{Features-statistics}.

\begin{table*}[htb]
    \begin{center}
         \caption{Features}
         \label{Features-statistics}
         \newcommand{\tabincell}[2]{\begin{tabular}{@{}#1@{}}#2\end{tabular}}
        \resizebox{0.95\textwidth}{!}{
           \begin{tabular}{l|l}
            \toprule
            \multicolumn{1}{c|}{\textbf{Information data}} &
            \multicolumn{1}{c}{\textbf{Features}}    \\
            \midrule
            \textbf{Entities}& companies, executives.\\
            \midrule
            \textbf{Relations}& explicit relations (industry category, supply chain, business partnership, investment), implicit relation.\\
            \midrule
            \textbf{Historical price}& opening price (op), closing price (cp), highest price (hp), lowest price (lp), trade volume (tv).\\
            \midrule
            \textbf{Media news}& positive media sentiment of a stock $Q(i)^{+}$, negative media sentiment of a stock  $Q(i)^{-}$, media sentiment divergence of a stock $D(i)$. \\
            \bottomrule
           \end{tabular}
         }
    \end{center}
\end{table*}


\subsection{Bi-typed Hybrid-relational Market Knowledge Graph (MKG)}
\label{section-mkgc}
\subsubsection{Bi-typed Entities}
Most existing SMP methods solely learn from company relationships in market \cite{Kim2019HATS,Ye2020Multi-Graph,Li2020Modeling}. 
{In fact, in most stock markets there are also significant associated executives for listed companies, with rich associative information about these companies \cite{cai2016price,jing2021online}. Hence, our constructed MKG contains not only company entities but also executive entities.}
The executive entities can act as the intermediary among companies to build the meta-relations involved in company entities (e.g. \textit{Company-Executive-Company (CEC)}, \textit{Company-Executive-Executive-Company (CEEC)}). 
For example, in Fig. \ref{figure-instance} we show the associated executives of the companies sampled from Chinese Stock Index. 
The stock spillover signals can pass from neighboring companies to a target company through the meta-relations established by their connected executives, such as Company 4-Executive C-Company 1; 
Company 2-Executive A$\stackrel{classmate}{\longleftrightarrow}$Executive B-Company 1.
In sum, the newly constructed MKG contains bi-typed entities, i.e. \textbf{\textit{listed companies}} and their \textbf{\textit{associated executives}}.



\subsubsection{Hybrid-relations}
Most existing methods only take the explicit relations among companies, such as \textit{industry category, supply chain, business partnership} and \textit{investment}, into consideration \cite{Chen2018Incorporating}. However, the limited explicit company relationships are always insufficient for market knowledge graph construction due to the complex nature of financial market. To solve the MKG incompleteness issue, here we propose an attribute-driven method to conduct MKG completion by inferring missing implicit correlative relation among stocks, which employs stocks attributions. 


Specifically, the attribute-driven implicit unobserved relation is calculated based on the features from both its historical prices and news information filtered by a normalized threshold. 
A single-layer feed-forward neural network is adopted to calculate the attention value $\alpha_{ij}^{t}$ between company $i$ and $j$ for inferring their implicit relation.

\begin{equation}
\alpha_{ij}^{t}=\mathop{\text{LeakyRelu}} \Big (\u ^{\top} [ \s_i ^t \parallel \s_j^t  ]\Big)  \ ,
\end{equation}
where 
$\s_{i}^{t}$ and $\s_{j}^{t}$ are fused market signals of $i$ and $j$, which are calculated by the Equation \ref{equaion-seq-learn} (Section \ref{section-se}). $\|$ denotes the concatenate operation. $\u$ denotes the learnable matrix and LeakyReLU is a nonlinearity activation function. Borrowing gate mechanism in \cite{Cheng2021Modeling}, we set an implicit relation between $i$ and $j$ if $\alpha_{ij}^{t} > \eta$. $\eta$ denotes a pre-defined threshold. 
In short, the constructed MKG contains hybrid-relations, i.e. \textbf{\textit{explicit relations}} and \textbf{\textit{implicit relation}}.
 

\subsection{Historical Price and Media News}
\label{section-msr}
\subsubsection{Technical Indicators}
Transactional data is the main manifestation of firms' intrinsic value and investors' expectations. We collect the daily stock price and volume data, including \textit{opening price (op), closing price (cp), highest price (hp), lowest price (lp), and trade volume (tv)}. 
In order to better compare and observe the fluctuation of stock price, the stock price is transferred to the return ratio, and the trade volume is transferred to the turnover ratio before being fed into our model. The return ratio is an index reflecting the level of stock return, the higher the return ration is; the better the profitability of the stock is. The turnover is the total value of stocks traded over a period of time; the higher the share turnover could indicate that the share haves good liquidity. $\p_i  \in \Rbb^5$ indicates the {technical indicators} of company $i$, as follows:
\begin{equation}
\label{equation-technical-indicator}
    \p_i = [op(i),\ cp(i),\  hp(i),\  lp(i),\  tv(i)]^{\top} \ .
\end{equation} 


\subsubsection{Sentiment Signals}
Modern behavioral finance theory \cite{Li2017Web} believes that investors are irrational, tending to be influenced by the opinions expressed in the media. Media sentiment reflects investors' expectations concerning the future of a company or the whole stock market, resulting in the fluctuations of stock price. To capture media sentiment signals, we extract the following characteristics: \textit{positive media sentiment, negative media sentiment} and \textit{media sentiment divergence} \cite{Li2020A}. They are denoted respectively as follows: 

\begin{equation}
\begin{aligned}
    Q(i)^{+} &=\frac{N(i)^{+}}{N(i)^{+}+N(i)^{-}} \ , \\
    Q(i)^{-} &=\frac{N(i)^{-}}{N(i)^{+}+N(i)^{-}}\ , \\
    D(i)& =\frac{N(i)^{+}-N(i)^{-}}{N(i)^{+}+N(i)^{-}} \ ,
\end{aligned}
\end{equation}
where $N(i)^{+}$ and $N(i)^{-}$ are the sum of the frequency of each positive and negative sentiment word found in the financial news articles of company $i$, respectively. $D(i)$ denotes the sentiment divergence. Since many negative sentiment words in the general sentiment dictionary no longer express negative emotional meanings
in the financial field, we resort to a finance-oriented sentiment dictionary created in previous study \cite{Li2016A}. $\q_i  \in \Rbb^3 $ indicates the news sentiment signals of company $i$.
\begin{equation}
\label{equation-sentiment-feature}
    \q_i =[Q(i)^{+},\ Q(i)^{-},\ D(i)]^{\top}\ .
\end{equation}

Note that we do not have everyday news for all companies since the randomness of the occurrence of media news. In order to make the technical indicators aligned with the media sentiment signals and keep pace with the real situation, the sentiment feature $\q_i$ of the firm $i$ is assigned to zero on the day when there are no any media news about it.

\section{Methodology}
\label{section-method}
In this section, we introduce the details of our proposed method. 
Figure \ref{figure-DanSmp-Model} gives an overview of the proposed framework. 
(I) First, the stock sequential embeddings are learned with historical price and media news via multi-modal feature fusion and sequential learning. 
(II) Second, a Dual Attention Networks is proposed to learn the stock relational embeddings based upon the constructed MKG. 
(III) Last, the combinations of sequential embeddings and relational embeddings are utilized to make stock prediction. 




\subsection{Learning Stock Sequential Embeddings}
\label{section-se}
\begin{figure}[t]
    \centering
    \includegraphics[width=1\textwidth]{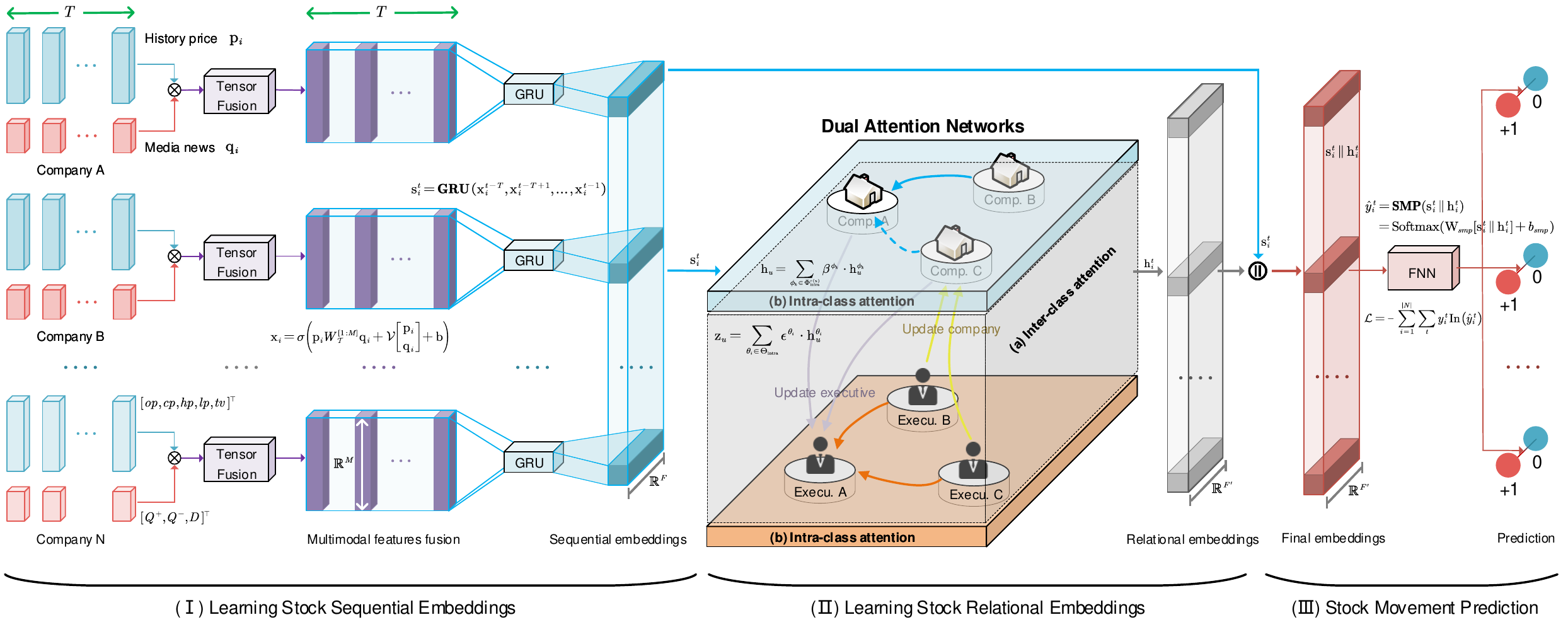}
    \caption{The overall framework of the proposed method. \textbf{(I) Learning Stock Sequential Embeddings} based on Tensor Fusion and GRU. Tensor Fusion is the Neural Tensor Network (NTN) to learn the fusion of the technical indicators vector $\p_i$ and media news sentiment features $\q_i$. The GRU is designed to learn the sequential embedding $\s_{i}^{t}$. \textbf{(II) Learning Stock Relational Embeddings} by a dual mechanism to model the mutual affects and inner interaction among the bi-typed entities (i.e. companies and executives) alternately, including: \textbf{(a)} inter-class attention, and \textbf{(b)} intra-class attention. The former aims to deal with the interaction between listed companies and their associated executives and the latter aims to learn the interaction among the same type of entities. \textbf{(III) Stock Movement Prediction} via Feed-forward Neural Network (FNN) with the learned firm embeddings.
    }
    \label{figure-DanSmp-Model}
\end{figure}

The stocks are influenced by multi-modal time-series market signals. 
Considering the strong temporal dynamics of stock markets, the historical state of the stock is useful for predicting its future trend. 
Due to the fact that the influence of market signals on the stock price would last for some time, we should consider the market signals in the past couple of days when predicting stock trend $\hat{y}_{i}^{t}$ of company $i$ at date $t$. 
We first capture the multimodal interactions of technical indicators and sentiment signals. 
We then feed the fused features into a one-layer GRU and take the last hidden state as the sequential embedding of stock $i$ which preserves the time dependency, as shown in Figure \ref{figure-DanSmp-Model}-I.

\subsubsection{Multimodal Features Fusion} 

To learn the fusion of the technical indicators vector $\p_i$ and media news sentiment features $\q_i$, we adopt a Neural Tensor Network (NTN) which replaces a standard linear neural network layer with a $M$-dimensional bilinear tensor layer that can directly relate the two features across multiple dimensions. 
The fused daily market signals\footnote{Here, the superscript $t$ is omitted for simplicity.} of stock $i$, $\x_{i} \in \Rbb^M$, are calculated by the tensor-based formulation as follows:
\begin{equation}
\x_{i}=\sigma \left ( \p_{i}W_{\Tcal}^{\left [ 1: M\right ]}\q_{i}+\Vcal\begin{bmatrix}
\p_{i}\\ \q_{i}
\end{bmatrix}+\b\right ) \ ,
\end{equation}
where $\sigma$ is an activation function, $W_{\Tcal}^{\left [ 1: M \right ]}\in \Rbb^{5\times 3 \times M}$ is a trainable tensor, $\Vcal \in \Rbb ^{ 8\times M}$ is the learned parameters matrix and $\b\in \Rbb^M$ is the bias vector. Three parameters are shared by all stocks.

\subsubsection{Sequential Learning} 
We feed the fused daily market signals in the past $T$ days into the GRU to learn {its} sequential embedding $\s_{i}^{t}$, as follows:
\begin{equation}
\label{equaion-seq-learn}
\s_{i}^{t}=\textbf{GRU}\left (\x_{i}^{t-T}, \x_{i}^{t-T+1}, \dots ,\x_{i}^{t-1} \right ) \ ,
\end{equation}
where $\s_{i}^{t}\in \Rbb^{F}$ denotes the last hidden state of GRU. ${F}$ is the hidden size of GRU.

\subsection{Learning Stock Relational Embeddings via Dual Attention Networks}
In real market, the stock fluctuation is partially affected by its related stocks which is known as momentum spillover effect in finance \cite{Ali2020Shared}. 
In this section, based upon our newly constructed bi-typed hybrid-relational MKG, we propose a Dual Attention Networks to learn the relational embeddings of stocks that represent their received spillover signals. Specifically, we employ a \textit{\textbf{dual mechanism}} to model the mutual affection and inner influence among the \textit{bi-typed} entities (i.e. companies and executives) alternately, including {inter-class interaction} and {intra-class interaction}, as shown in Figure \ref{figure-DanSmp-Model}-II.

\subsubsection{Inter-class Attention Networks}
The inter-class attention aims to deal with the interaction between listed companies and their associated executives, as shown in Figure \ref{figure-DanSmp-Model} (II-a). Since they are different types of entities, their features usually lie in different space. Hence, we first project their embeddings into a common spaces. Specifically, for a company entity $u \in \Ecal_1$ with type $\tau(u)$ and an executive entity $v \in \Ecal_2$ with type $\tau(v)$, we design two type-specific matrices
$\W^{\tau(\cdot)}$ 
to map their features $\h_{u},\h_v$ into a common space.
\begin{equation}
\begin{array}{l}
\begin{aligned}
    \label{equation-att-node}
     \h_u'&=\W^{\tau(u)}\h_u \ , \quad \\
     \h_v'&=\W^{\tau(v)}\h_v \ ,
\end{aligned}
\end{array}
\end{equation}
where $\h_u'\in \Rbb^{F'}$ and $\h_u\in \Rbb^{F}$ denote the original and transformed features of the entity $u$, respectively. 
$\Ecal_1$ and $\Ecal_2$ are the sets of listed companies and the executives, respectively.
Here, the original vectors of company entities ($\h_u,\h_v$) are initialized by learned sequential embeddings ($\s_u,\s_v$) learned in Section \ref{section-se}, which can bring rich semantic information in downstream learning. The initial features of executives are then simply an average of the features of the companies which they work for.

We assume that the target company entity $u$ connects with other executives via a relation ${\theta_i \in \Theta_{\text{inter} }}$ which denotes the set of inter-class relations, so the neighboring executives of a company $u$ with relation ${\theta_i}$ can be defined as $\Ncal_\text{inter}^{\theta_i}(u)$. For entity $u$, different types of inter-class relations contribute different semantics to its embeddings, and so do different entities with the same relation. Hence, we employ attention mechanism here in entity-level and relation-level to hierarchically aggregate signals from other types of neighbors to target entity $u$.
%

We first design an entity-level attention to learn the importance of entities within a same relation. 
Then, to learn the importance $e_{u\upsilon}^{\theta_i}$ which means how important an executive $v$ for a company $u$ under a specific relation $\theta_i$, we perform {self-attention} \cite{Vaswani2017Attention} on the entities as follows:
\begin{equation}
\begin{array}{l}
\begin{aligned}
    \label{equation-att-node}
     e_{uv}^{\theta_i}  &=  att_{node}(\h_u',\h_v';\theta_i) \\
     &= \text{LeakyRelu}(\a_{\theta_i}^\top \cdot [ \h_u' \|  \h_v']) \ ,
\end{aligned}
\end{array}
\end{equation}
where $\h_u'$ and $\h_\upsilon'$ are the transformed representations of the node $u$ and $\upsilon$. $\a_{\theta_i}\in \Rbb^{2F'}$ is a trainable weight vector. $\|$ denotes the concatenate operation. LeakyReLU is a nonlinearity activation function. 
To make $e_{u\upsilon}^{\theta_i}$ comparable over different entities, we normalize it using the softmax function.
\begin{equation}
    \gamma_{uv}^{\theta_i} =\text{softmax}_\upsilon (e_{uv}^{\theta_i})= \frac{\exp{(e_{uv}^{\theta_i}})}{\sum\limits_{\bar{v} \in \Ncal_\text{inter}^{\theta_i}(u) }\exp{(e_{u\bar{v}}^{\theta_i})}} \ ,
\end{equation}
where $\gamma_{uv}^{\theta_i}$ denotes the attention value of entity $v$ with relation $\theta_i$ to entity $u$. $\Ncal_\text{inter}^{\theta_i}(u)$ denotes the specific relation-based neighbors with the different type.
%
We apply entity-level attention to fuse inter-class neighbors with a specific relation $\theta_i$:
\begin{equation}
    \label{equation-node-level-aggregation}
    \h_u^{\theta_i} = \sigma \Big(\sum_{v \in \Ncal_\text{inter}^{\theta_i}(u)}\gamma_{uv}^{\theta_i} \cdot \h_v' \Big ) \ ,
\end{equation}
where $\sigma$ is a nonlinear activation, and $\h_v'$ is the projected feature of entity $v$.  



Once we learned all relation embeddings $\{\h_u^{\theta_i}\}$, we utilize relation-level attention to fuse them together to obtain the inter-class relational embedding $z_u$ for entity $u$.
We first calculate the importance of each relation $w^{\theta_i}$ as follows:
\begin{equation}
   w^{\theta_i}=\frac{1}{|\Ecal_1|}\sum\limits_{u\in \Ecal_1} \q^{\tau(u)} \cdot \h_u^{\theta_i} + \frac{1}{|\Ecal_2|}\sum\limits_{v\in \Ecal_2} \q^{\tau(v)} \cdot \h_v^{\theta_i} \ ,
\end{equation}

\begin{equation}
    \epsilon^{\theta_i} = \frac{\exp{(w^{\theta_i})}}{\sum\limits_{\theta_j \in \Theta_{\text{inter} }}\exp{(w^{\theta_j})}} \ ,
\end{equation}
where $\q^{\tau(\cdot)}\in \Rbb^{F'\times 1}$ is learnable parameter. We fuse all relation embeddings to obtain the inter-class relational embedding $\z_u\in \Rbb^{F'}$ of entity u. 
\begin{equation}
    \label{equation-node-level-aggregation}
    \z_u  = \sum_{\theta_i \in \Theta_{\text{intra} }} \epsilon^{\theta_i }\cdot \h_u^{\theta_i} \ .
\end{equation}
In inter-class attention, the aggregation of different entities' embedding are seamlessly integrated, and they are mingled and interactively affected each other in nature, as shown in Figure \ref{figure-DanSmp-Model} (II-a).

\subsubsection{Intra-class Attention Networks}
The intra-class attention aims to learn the interaction among the same type of entities, as shown in Figure \ref{figure-DanSmp-Model} (II-b).
Specifically, given a relation $\phi_k \in \Phi_\text{intra}^{\tau(u)}$ that starts from entity $u$, we can get the intra-class relation based neighbors $\Ncal_\text{intra}^{\phi_k}(u)$.  $\Phi_\text{intra}^{\tau(u)}$ indicates the set of all intra-class relations of $u$. For instance, as shown in Figure \ref{figure-instance}, Company 5 is a neighbor of Company 3 based on an implicit relation, and Company 4 is a neighbor of Company 1 based on meta-relation \textit{CEC}. Each intra-class relation represents one semantic interaction, and we apply relation-specific attention to encode this characteristic. We first calculate the attention value of entity $\tilde{u}$ with relation $\phi_k$ to entity $u$ as follows:




\begin{equation}
    \alpha_{u\tilde{u}}^{\phi_k} =
    \frac{\exp{(\text{LeakyRelu}(\a_{\phi_k}^\top \cdot [\W \z_u \| \W \z_{\tilde{u}}]))}}{\sum\limits_{u' \in \Ncal_\text{intra}^{\phi_k}(u) }\exp{(\text{LeakyRelu}(\a_{\phi_k}^\top \cdot [\W \z_u \| \W \z_{u'}]))}} \ ,
\end{equation}
where $\z_u$ and $\z_{\tilde{u}}$ are output representations of the inter-class attention, respectively. $\W\in \Rbb^{F'\times F'}$ is a trainable weight matrix which is shared to every node of the same type.
$\a_{\phi_k}\in \Rbb^{2F'}$ is the node-level attention weight vector for relation $\phi_k$.
$\Ncal_\text{intra}^{\phi_k}(u)$ denotes the intra-class neighbors of $u$ under relation $\phi_k$. 
The embedding $\h_u^{\phi_k}$ of entity $u$ for the given relation $\phi_k$ is calculated as follows.
\begin{equation}
    \label{equation-node-level-aggregation}
    \h_u^{\phi_k}  = \sigma \Big(\sum_{\tilde{u} \in \Ncal_\text{intra}^{\phi_k}(u)}\alpha_{u\tilde{u}}^{\phi_k} \cdot \W\z_{\tilde{u}} \Big ) \ ,
\end{equation}
where $\sigma$ is a non-linear activation. In total, we can get $|\Phi_\text{intra}^{\tau(u)}|$ embeddings for entity $u$. Then, we conduct relation-level attentions to fuse them into the relational embedding $\h_u\in \Rbb^{F'}$:

\begin{equation}
    \label{equation-node-level-aggregation}
    \h_u  = \sum_{\phi_k \in \Phi_{\text{intra}}^{\tau(u)}} \beta^{\phi_k} \cdot \h_u^{\phi_k} \ ,
\end{equation}
where $\Phi_\text{intra}^{\tau(u)}$ denotes the set of all intra-class relationships of entity $u$. $\beta^{\phi_k}$ denotes the importance of intra-class relation $\phi_k$, which is calculated as follows:
\begin{equation}
\begin{array}{l}
\begin{aligned}
   \g^{\phi_k}&=\frac{1}{|\Ecal_1|}\sum\limits_{u\in \Ecal_1} \q^{\tau(u)} \cdot \h_u^{\phi_k} \ ,\\
   \beta^{\phi_k} &= \frac{\exp{(\g^{\phi_k})}}{{\sum\limits_{\phi_l \in \Phi_\text{intra}^{\tau(u)} }}
    \exp{(\g^{\phi_l})}} \ .
\end{aligned}
\end{array}
\end{equation}
where $\q^{\tau(u)}\in \Rbb^{F'}$ is a learnable parameter.

\subsection{SMP with Stock Final Embeddings}
Finally, the stock final embeddings by combining learned sequential embeddings and relational embeddings are utilized to make stock prediction by a dense layer feed-forward neural network (FNN) and a softmax function, as shown in Figure \ref{figure-DanSmp-Model}-III.
\begin{equation}
\begin{aligned}
    \hat{y}_{i}^{t}&=\textbf{SMP} \Big(\s_{i}^{t} \parallel \h_{i}^{t} \Big) \\ 
    &= \text{Softmax} \Big (\W_{smp} [ \s_{i}^{t} \parallel \h_{i}^{t} ]+ b_{smp} \Big )  \ ,
\end{aligned}
\end{equation}
where $\W_{smp}$ is a trainable weight matrix, and $b_{smp}$ is the bias vector. 
We leverage the Adam algorithm \cite{Kingma2014Adam} for optimization by minimizing the cross entropy loss function $\Lcal$.

\begin{equation}
\Lcal=-\sum_{i=1}^{\left | N\right |}\sum_t {y}_{i}^t ln\left ( \hat{y}_{i}^t\right ) \ ,
\end{equation}
where ${y}_{i}^t$ and $\hat{y}_{i}^t$ represent the ground truth and predict stock trend of stock $i$ at $t$ day, respectively. $\left | N\right |$ is the total number of stocks.

\section{Experiments}
\label{section-experiments}

In this section, we present our experiments, mainly focusing on the following research questions:

$\bullet$ RQ1: Can our model achieve better performance than the state-of-the-art stock prediction methods?

$\bullet$ RQ2: Can our model achieve a higher investment return and lower risk in the investment simulation on real-world datasets?

$\bullet$ RQ3: How is the effectiveness of different components in our model?

$\bullet$ RQ4: Are all firm relations equally important for SMP? How do different parameters influence our model’s performance?

In the following, we first present the experimental settings and then answer these research questions by analyzing the experimental results.

\subsection{Experimental Settings} 
\subsubsection{Data Collection}
Since no existing stock prediction benchmark datasets can satisfy our need to evaluate the effectiveness of our method, we collect public available data about the stocks from the famous China Securities Index (CSI) and construct two new datasets. 
We name them \textbf{CSI100E} and \textbf{CSI300E}
with different number of listed companies, respectively. 
185 stocks in CSI300E index without missing transaction data and having at least 60 related news articles during the selected period are kept. Similarly, 73 stocks in CSI100E index are kept.
First, we get historical price of stocks\footnote{We collect daily stock price and volume data from \url{https://www.wind.com.cn/}} from November 21, 2017 to December 31, 2019 which include 516 transaction days.  
Second, we collect web news published in the same period from four financial mainstream sites, including \textit{Sina\footnote{\url{http://www.sina.com}}, Hexun\footnote{\url{http://www.hexun.com}}, Sohu\footnote{\url{http://www.sohu.com}}} and \textit{Eastmoney}\footnote{\url{http://www.eastmoney.com}}. 
Last, we collect four types of company relations\footnote{We collect four types of company relations by a publicly available API tushare: \url{https://tushare.pro/}.} and the connections of executives\footnote{We collect executives relationships from : \url{http://www.51ifind.com/}.} for CSI300E and CSI100E.
The basic statistics of the datasets are summarized in Table \ref{data-stcs}.
The usage details of the multimodal market signals are described in Section \ref{section-msr}. 


\begin{table}[t]
\caption{Statistics of datasets.}
\label{data-stcs}
\centering
\begin{tabular}[t]{l||c|c}
\toprule
{ }  &\textbf{CSI100E} & \textbf{CSI300E}\\
\midrule
\textbf{\textit{\#Companies(Nodes)}}  & 73 & 185  \\
\textbf{\textit{\#Executives(Nodes)}}  & 163& 275  \\
\midrule
\textbf{\textit{\#Investment(Edges)}}  & 7& 44  \\
\textbf{\textit{\#Industry category(Edges)}}  & 272& 1043  \\
\textbf{\textit{\#Supply chain(Edges)}}  & 27& 37  \\
\textbf{\textit{\#Business partnership(Edges)}}  & 98& 328  \\
\textbf{\textit{\#Implicit relation(Edges)}}  & \textit{dynamic}& \textit{dynamic}  \\
\textbf{\textit{\#meta-relation CEC}}  &  18&  42 \\
\textbf{\textit{\#meta-relation CEEC}}  &  134&  252 \\
\midrule
\textbf{\textit{\#Classmate(Edges) }}  & 338& 592  \\
\textbf{\textit{\#Colleague(Edges)}}  & 953& 2224  \\
\midrule
\textbf{\textit{\#Management(Edges)}}  & 166& 275  \\
\textbf{\textit{\#Investment(Edges)}}  & 1& 8  \\
\midrule
\textbf{\textit{\bf\#Train Period}}  & 21/11/2017-05/08/2019 & 21/11/2017-05/08/2019  \\
\textbf{\textit{\bf\#Valid Period}}  & 06/08/2019-22/10/2019 & 06/08/2019-22/10/2019  \\
\textbf{\textit{\bf\#Test Period}}  & 23/10/2019-31/12/2019 & 23/10/2019-31/12/2019  \\
\bottomrule
\end{tabular}
\end{table}
  
\subsubsection{Evaluation Protocols}
SMP is usually treated as a binary classification problem. 
If the closing price of a stock $i$ is higher than its opening price at day $t$, the stock movement trend is defined as “upward” $\left ( y_{i}^{t}=1\right )$, otherwise as “downward” $\left ( y_{i}^{t}=0\right )$. According to statistics, there are 46.7$\%$ “upward” {stocks} and 53.3$\%$ “downward” ones in CSI100E, and 47.8$\%$ “upward” and 52.2$\%$ “downward” {stocks} in CSI300E. Hence, the datasets are roughly balanced. 

Some indicators \cite{sousa2019bert,Ye2020Multi-Graph} are selected to demonstrate the effectiveness of the proposed method, i.e. Directional Accuracy (DA), Precision,  (AUC), Recall, F1-score. 
We use the \textbf{Directional Accuracy (DA)} and \textbf{AUC} (the area under the precision-recall curve) 
to evaluate classification performance in our experiments, 
which are widely adopted in previous works \cite{Li2020A,Cheng2021Modeling}. 
Similar to \cite{Sawhney2020Spatiotemporal,Sawhney2021Stock}, 
to evaluate \textsc{DanSmp}'s applicability to real-world trading, we assess its profitability on CSI100E and CSI300E using metrics: cumulative investment return rate (\textbf{IRR}) and \textbf{Sharpe Ratio} \cite{sharpe1994sharpe}.
Similar to previous method \cite{Li2020A,Cheng2021Modeling}, we use the market signals of the past $T$ trading days (also called lookback window size) to predict stock movement on $t^{th}$ day. The DA, IRR and SR are defined as follows:
\begin{equation}
\begin{aligned}
    DA &=\frac{n}{N} \ , \\
    IRR^{t} &=\sum_{i\in S^{t-1}}\frac{p_{i}^{t}-p_{i}^{t-1}}{p_{i}^{t-1}} \ ,  \\
    SR_{a}& =\frac{E\left [ R_{a}-R_{f}\right ]}{std\left [ R_{a}-R_{f}\right ]} \ ,
\end{aligned}
\end{equation}
where $n$ is the number of predictions, which witness the same direction of stock movements for the predicted trend and the actual stock trend and $N$ is the total number of predictions. $S^{t-1}$ denotes the set of stocks on day $t-1$, and  $p_{i}^{t}$ is the price of stock i at day $t$. $R_{a}$ denotes an asset return and $R_{f}$ is the risk-free rate. In this study, the risk-free rate is set as the one-year deposit interest rate of the People's Bank of China in 2019, i.e. $R_{f}=1.5\%$.

Note that, to ensure the robustness of the evaluation, we repeat the testing procedure 10 times with different initialization for all the experimental results and the average performance is reported as the final model result.

\subsubsection{Parameter Settings}
All trainable parameters vectors and matrices are initialized using the Glorot initialization \cite{glorot2010understanding}.
In our \textsc{DanSmp}, we set the lookback window size $T$ among [10, 15, 20, ... , 40]. We search the learning rate from
[0.00005, 0.0001, 0.00015, ... , 0.002]. The slice size of NTN $M$ and attention layer hidden size ${F}'$ are determined in [5,10,15, ... ,50] and [10, 11, 12, ... , 50], respectively. The GRU hidden size $F$ is set from [20, 22, 24, ... , 100]. In our model, all hyperparameters were optimized with the validation set, and Table \ref{table-hyperparameters} shows the hyper-parameter settings of our method. The proposed
\textsc{DanSmp} is implemented with PyTorch\footnote{\url{https://pytorch.org/}.} 
and PyTorch Geometric\footnote{\url{https://pytorch-geometric.readthedocs.io/en/latest/}.}, 
and each training process costs 1.5hrs averagely using a GTX 1080 GPU. To prevent overfitting, we use early stopping based on AUC (the area under the precision-recall curve) over the validation set.

\begin{table}[htb]
\caption{The hyper-parameter settings on two datasets.}
\label{table-hyperparameters}
\newcommand{\tabincell}[2]{\begin{tabular}{@{}#1@{}}#2\end{tabular}}
\centering
\begin{tabular}[t]{l||c c}
\toprule
\bf Parameter & \bf CSI100E & \bf CSI300E\\
\midrule
Lookback window size $T$& 20 & 20 \\ 
The slice size of NTN $M$ & 10 & 10\\
Attention layer hidden size ${F}'$ & 39 & 22\\
GRU hidden size $F$& 78 & 44 \\
Learing rate & 0.0008 & 0.00085\\
Implicit relation threshold $\eta$ & 0.0054 & 0.0052 \\
Maximum number of epochs  & 400 & 400 \\
\bottomrule
\end{tabular}
\end{table}


\subsubsection{Baselines}

To demonstrate the effectiveness of our proposed model \textsc{DanSmp}, we compare the results with the following baselines. 

$\bullet$ LSTM \cite{Hochreiter1997Long}: a typical RNN model that has promising performance on time-series data. In the evaluation, two-layer LSTM networks are implemented. \todo{}

$\bullet$ GRU \cite{Cho2014Learning}: a simpler RNN that achieves similar performance with LSTM. In the comparison, two-layer GRU networks are implemented.


$\bullet$ GCN \cite{Kipf2017Semi-supervised}: It performs graph convolutions to linearly aggregate the attributes of the neighbor nodes. In this study, two-layer GCN network was implemented.  \todo{}

$\bullet$ GAT \cite{Velickovic2018Graph}: It introduces attention mechanism which assigns different importance to the neighbors adaptively.  Two-layer GAT networks are implemented.

$\bullet$ RGCN \cite{Schlichtkrull2018Modeling}: It designs specialized mapping matrices for each relations.  Two-layer RGCN network was implemented.

$\bullet$ HGT \cite{Hu2020Heterogeneous}: It uses transformer architecture to capture features of different nodes based on type-specific transformation matrices.


$\bullet$ MAN-SF \cite{Sawhney2020Deep}: It fuses chaotic temporal signals from financial data, social media and stock relations in a hierarchical fashion to predict future stock movement.

$\bullet$ STHAN-SR \cite{Sawhney2021Stock}: It uses hypergraph and temporal Hawkes attention mechanism to rank stocks with only historical price data and explicit firm relations. We only need to slightly modify the objective function of MAN-SF to predict future stock movement.

$\bullet$ AD-GAT \cite{Cheng2021Modeling}: a SOTA method to use an attribute-driven graph attention network
to capture attribute-sensitive momentum spillover of stocks, which can modeing market information space with feature interaction to further improve stock movement prediction.


These baselines cover different model characters. 
Specifically, the sequential-based LSTM \cite{Hochreiter1997Long} and GRU \cite{Cho2014Learning} can capture the time dependency of stock data, and the fused market signals were used as the input to the LSTM and GRU model. 
The homogeneous GNNS-based GCN \cite{Kipf2017Semi-supervised}, GAT \cite{Velickovic2018Graph}, RGCN \cite{Schlichtkrull2018Modeling}, HGT \cite{Hu2020Heterogeneous}, MAN-SF \cite{Sawhney2020Deep}, STHAN-SR \cite{Sawhney2021Stock} and AD-GAT \cite{Cheng2021Modeling} can capture the influence of related stocks based on the fused market signals and simple firm relations. 
Note that, for fair comparison, we do not select the methods that are incapable of dealing with all fused multi-modal market signals (i.e. historical price, media news and stock relations) as baselines. 

\begin{table}[t]
    \centering
    \caption{Stock prediction results of different models.}
    \label{table-model-comparison}
    \newcommand{\tabincell}[2]{\begin{tabular}{@{}#1@{}}#2\end{tabular}}
    \centering
    \begin{tabular}{l||cc|cc}
    \toprule    
    \multirow{2}*{\bf{Methods} } &\multicolumn{2}{c|}{\bf CSI100E} &\multicolumn{2}{|c}{\bf CSI300E}\\
       &\textbf{Accuracy}&\textbf{AUC}&\textbf{Accuracy} &\textbf{AUC} \\
    \midrule
     \midrule
        LSTM \cite{Hochreiter1997Long} & 51.14 & 51.33&51.78  &52.24  \\
        GRU \cite{Cho2014Learning} & 51.66 & 51.46&51.11  &52.30  \\
    \midrule
    GCN \cite{Kipf2017Semi-supervised} & 51.58 &52.18&51.68 & 51.81   \\
    GAT \cite{Velickovic2018Graph} & 52.17 &  52.78&51.40 & 52.24 \\
    RGCN \cite{Schlichtkrull2018Modeling} & 52.33 &  52.69&51.79 & 52.59 \\
    HGT \cite{Hu2020Heterogeneous} & 53.01 &52.51&51.70 & 52.19  \\
    MAN-SF \cite{Sawhney2020Deep}& 52.86 & 52.23 &51.91 & 52.48\\
    STHAN-SR \cite{Sawhney2021Stock}& 52.78 & 53.05 & \underline{52.89}& 53.48\\
    AD-GAT \cite{Cheng2021Modeling} &\underline{54.56} & \underline{55.46}&52.63 &\underline{54.29}\\
    \midrule
    \textbf{\textsc{DanSmp} (ours)} & \textbf{57.75}  & \textbf{60.78} & \textbf{55.79}& \textbf{59.36} \\
    \bottomrule
    \end{tabular}
\end{table}
\subsection{Experimental Results and Analysis (RQ1)}
Table \ref{table-model-comparison} shows the evaluation results of the two datasets against nine state-of-the-art (SOTA) baselines, from which we observe that our proposed method outperforms all baselines for stock movement prediction in terms of all metrics on CSI100E and CSI300E. 
It confirms the capability of our method in modeling the comprehensive market signal representations via dual attention networks.

\paragraph{\textbf{Analysis.}}

(1) The LSTM and GRU, which only consider historical prices and media news, perform largely worse than our method. The results indicate that the relation datas contribute to stock movement prediction and the proposed method can take full advantage of the relational information in MKG to improve performance.  
(2) The graph-based methods, such as GCN and GAT, are homogeneous GNNs which are incapable of modeling heterogeneous market graph. Although being able to model multi-relational graph, RGCN can not sufficiently encode bi-typed heterogeneous graph become of the fact that it ignores the heterogeneity of node attributes and calculates the importance of neighbors within the same relation based on predefined constants.
HGT focuses on handling web-scale heterogeneous graphs via graph sampling strategy, which thus is prone to overfitting when dealing with relative sparse MKG. HGT can not learns multi-level representation by sufficiently utilize interactions between two types of nodes.
We believe that is the reason they perform worse than our model \textsc{DanSmp} which is pertinently designed to model bi-typed hybrid-relational MKG.
(3) The proposed \textsc{DanSmp} consistently outperforms three other SMP competitors, including AD-GAT, STHAN-SR and MAN-SF.
Specifically, it exceeds the second place by approximately 3.19$\%$ and 5.32$\%$ in terms of Accuracy and AUC in CSI100E, and 3.16$\%$ and 5.07$\%$ in CSI300E. The results clearly demonstrate the effectiveness of \textsc{DanSmp} and the explicit relation and executives relation are meaningful for stock movement prediction. 



\begin{table}[t]
    \centering
    \caption{Profitability of all methods in back-testing.}
    \label{table-as-four}
    \newcommand{\tabincell}[2]{\begin{tabular}{@{}#1@{}}#2\end{tabular}}
    \centering
    \begin{tabular}{l||rr|rr}
    \toprule    
    \multirow{2}*{\bf{Methods} } &\multicolumn{2}{c|}{\bf CSI100E} &\multicolumn{2}{|c}{\bf CSI300E}\\
       &\textbf{IRR}&\textbf{SR}&\textbf{IRR} &\textbf{SR} \\
    \midrule
     \midrule
        LSTM \cite{Hochreiter1997Long} &-4.57\% & -2.1713&-0.38\%  &-0.326  \\
        GRU\cite{Cho2014Learning} & -2.55\% & -1.053&-3.73\%  &-1.197  \\
    \midrule
    GCN \cite{Kipf2017Semi-supervised} & 1.59\% &0.719&3.55\% & 1.873   \\
    GAT \cite{Velickovic2018Graph} & 0.3\% &  0.050&-1.82\% & -1.121 \\
    RGCN \cite{Schlichtkrull2018Modeling}& 6.41\% &  3.789&-3.64\% & -1.905 \\
    HGT\cite{Hu2020Heterogeneous} & 2.54\% &1.716&0.36\% & 0.076  \\
    MAN-SF\cite{Sawhney2020Deep}& -2.91\% & -1.590 &1.38\% & 0.604\\
    STHAN-SR\cite{Sawhney2021Stock}& -0.12\% & -0.092 & 5.41\%& 1.565\\
    AD-GAT \cite{Cheng2021Modeling} &2.34\% & 1.190&15.12\% &4.081\\
    \midrule
    \textbf{\textsc{DanSmp} (ours)} & \textbf{10.18\%}  & \textbf{4.112} & \textbf{16.97\%}& \textbf{4.628} \\
    \bottomrule
    \end{tabular}
\end{table}

\begin{figure}[htb]
    \centering
    \includegraphics[width=0.8\textwidth]{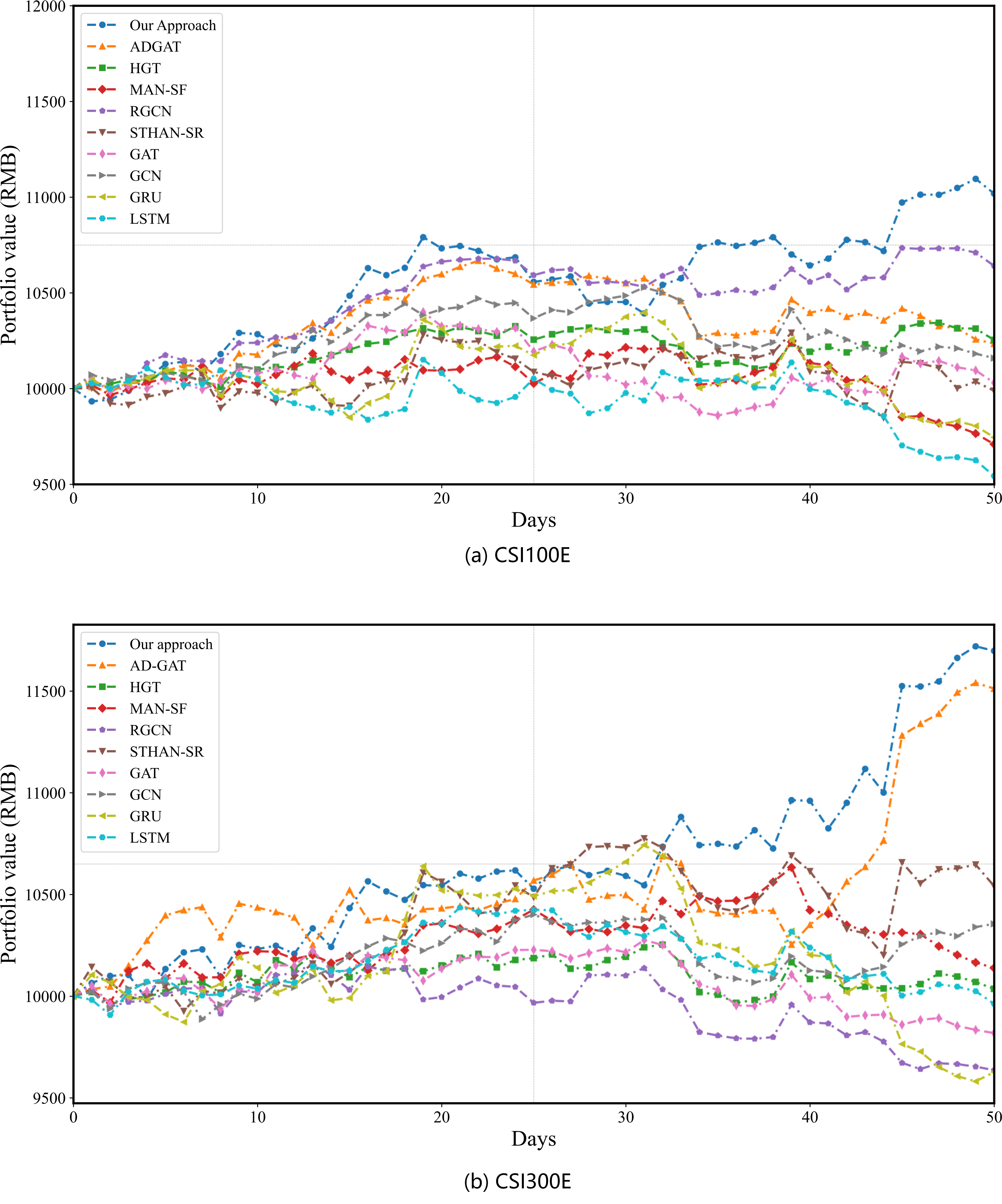}\\
    \caption{Profitability analysis on CSI100E and CSI300E.}
    \label{figure-investment-simulation}
\end{figure}

\subsection{Investing simulation (RQ2)}
To test whether 
a model can make a profit, we set up a back-testing via simulating the stock investment in CSI100E and CSI300E over the test period, during which the CSI100 and CSI300 index increased by 4.20\% and 5.14\% (from 4144.05 to 4317.93 and 3896.31 to 4096.58), respectively. Specifically, the top-15 stocks with the highest predicted ranking score in each model are bought and held for one day. We choose RMB 10,000 as the investment budget, and take into account a transaction cost of 0.03\% when calculating the investment return rate, which is in accordance with the stock market practice in China. The cumulative profit will be invested into the next trading day.
From Table \ref{table-as-four} and Figure \ref{figure-investment-simulation}, we can find that \textsc{DanSmp} achieves approximately stable and continuous positive returns throughout the back-testing. Particularly, the advantage of \textsc{DanSmp} over all baselines mainly lies in its superior performance when the stock market is in a bear stage. The proposed \textsc{DanSmp} achieves significantly higher returns than all baselines with the cumulative rate of 10.18\% and 16.97\% in CSI100E and CSI300E. In addition, \textsc{DanSmp} results in a more desirable risk-adjusted return with Sharpe Ratio of 4.113 and 4.628 in CSI100E and CSI300E, respectively. These results further demonstrate the superiority of the proposed method in terms of the trade-off between the return and the risk.

\subsection{Ablation Study (RQ3)}
To examine the usefulness of each component in \textsc{DanSmp}, we conduct ablation studies on CSI100E and CSI300E. We design four variants: 
(1) \textbf{\textsc{DanSmp} w/o executives}, which deletes the executive entities. MKG is degraded into a simple uni-type knowledge graph.
(2) \textbf{\textsc{DanSmp} w/o implicit relation}, which removes the implicit relation when we model the stock momentum spillover effect.
(3) \textbf{\textsc{DanSmp} w/o explicit relation}, which deletes the explicit relations and only use the implicit relation to predict stock movement.
(4) \textbf{\textsc{DanSmp} w/o dual}, which replaces the dual attention module by conventional attention mechanism and does not distinguish the node intra-class and inter-class relation. 

From Table \ref{table-as-three}, we observe that 
removing any component of \textsc{DanSmp} would lead to worse results. 
The effects of the four components vary in different datasets, but all of them contribute to improving the prediction performance.  
Specifically, removing executives relations and implicit relations leads to the most performance drop, compared to the other two, which means a company can influence the share price of other companies through interactions between executives. In contrast, using the conventional attention mechanism produces the least performance drop. Compared with conventional attention mechanism, the dual attention module enables \textsc{DanSmp} to adaptively select more important nodes and relations. 
This finding further proves that the proposed \textsc{DanSmp} fully leverages bi-typed hybrid-relational information in MKG via dual mechanism for better stock prediction.

\begin{table}[t]
    \centering
    \caption{The ablation study over \textsc{DanSmp}.}
    \label{table-as-three}
    \newcommand{\tabincell}[2]{\begin{tabular}{@{}#1@{}}#2\end{tabular}}
    \centering
    \begin{tabular}{l||cc|cc}
    \toprule    
    \multirow{2}*{\bf{Variants} } &\multicolumn{2}{c|}{\bf CSI100E} &\multicolumn{2}{|c}{\bf CSI300E}\\
       &\textbf{Accuracy}&\textbf{AUC}&\textbf{Accuracy} &\textbf{AUC} \\
   \midrule

    \textsc{DanSmp} & \textbf{57.75} &\textbf{60.78}& \textbf{55.79}& \textbf{59.36} \\
    \midrule
    \  w/o executives &52.71 & 52.62&53.80&55.88 \\
    \  w/o implicit rel. &53.52 & 54.38&52.13&53.37 \\
    \  w/o explicit rel. &55.12 & 57.05&54.10&55.49 \\
    \midrule
    \  w/o dual &56.12 & 58.60&55.43&57.85 \\

     \bottomrule
    \end{tabular}
\end{table}

\begin{figure}[t]
    \centering
    \includegraphics[width=0.9\textwidth]{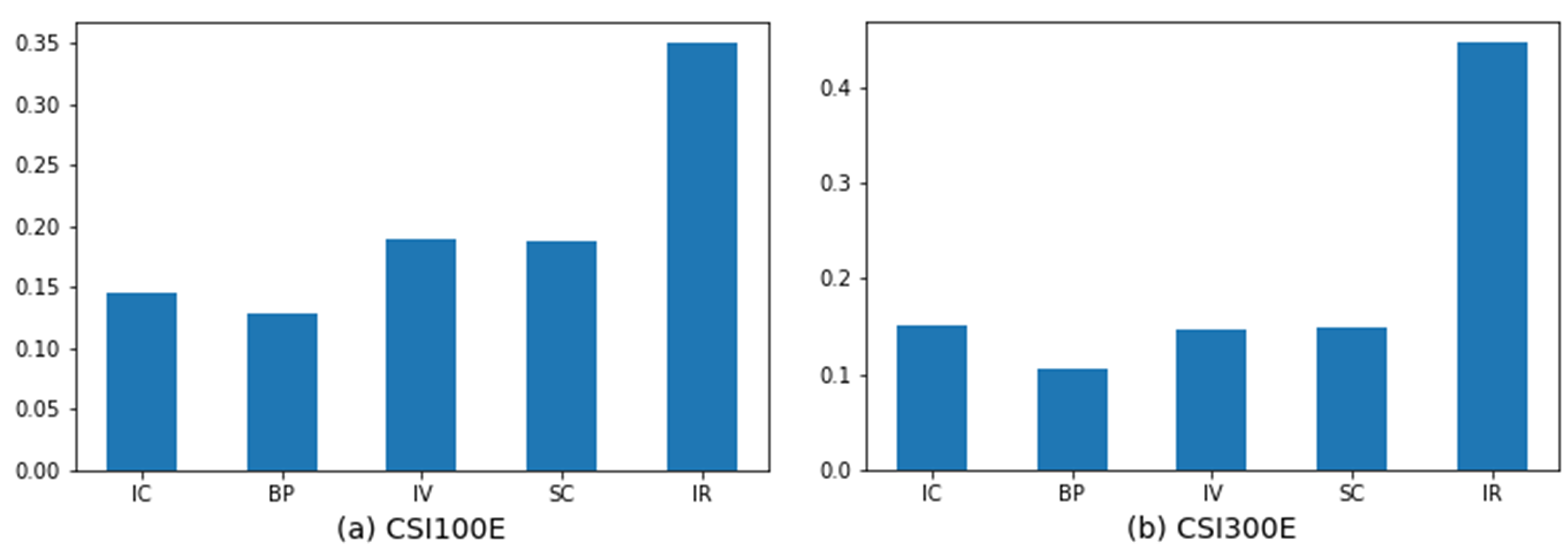}\\
    \caption{The presentation of the learned attention scores of \textsc{DanSmp} on CSI100E and CSI300E. Here, IC denotes the industry category; BP stands for the business partnership; IV denotes the investment; SC is the supply chain; IR denotes the implicit relation.}
    \label{figure-attention-scores}
\end{figure}

\begin{figure}[htb]
    \centering
    \includegraphics[width=0.95\textwidth]{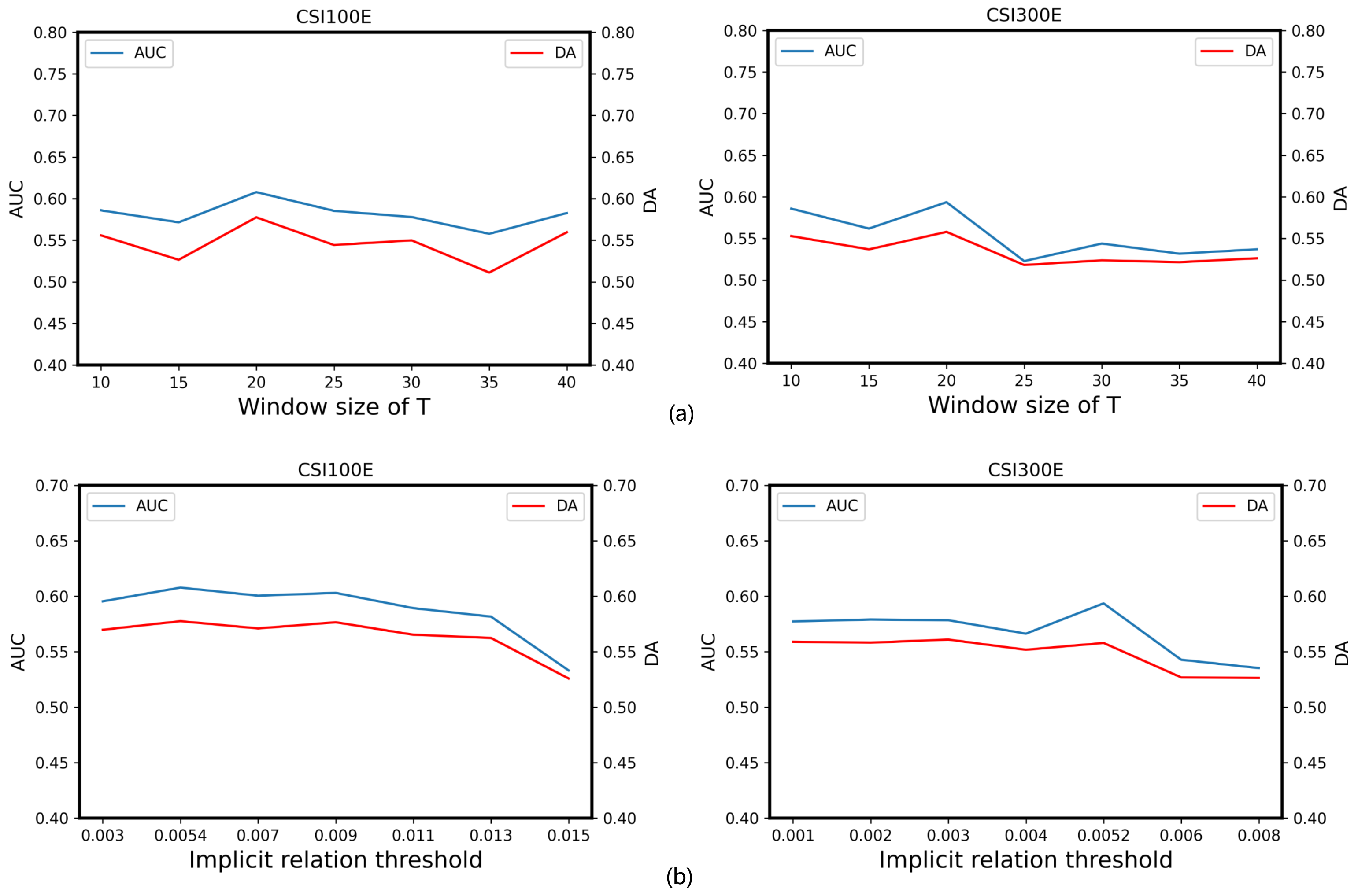}
    \caption{Sensitivity to parameters $T$ and $\eta$.}
    \label{figure-parameter-analysis-1}
\end{figure}

\subsection{Analysis of Firm Relation (RQ4)}
To investigate the impact of using different types of relations for stock prediction, we show the learned attention scores of our model in Figure \ref{figure-attention-scores}. The attention score is learned parameter for different firm relations. Some main findings are as follows: 
(1) We can observe Figure \ref{figure-attention-scores} that the learned attention score of implicit relation gains more weight than other relations, and the implicit relation which contains a lot of valuable information proved to be helpful for stock movement prediction. 
(2) The industry category, supply chain and investment get almost the same attention scores, and those can improve the
performance of model. 
(3) Compared with other relations, the business partnership has the lowest score. Although the number of business partnership relation is greater than that of investment and supply chain, the relatively dense business partnership relation may carry some noise, which adds irrelevant information to the representations of target nodes. The results further demonstrate the necessity of considering the implicit relation in modeling the momentum spillover effect. In addition, our model can adaptively weight important company relations to obtain better representations of target nodes, which can improve the performance of the model for stock movement prediction.

\subsection{Parameter Sensitivity Analysis (RQ4)}
We also investigate on the sensitivity analysis of two parameters in \textsc{DanSmp}. We report the results of \textsc{DanSmp}  under different parameter settings on CSI100E and CSI300E and experimental results are shown in Figure \ref{figure-parameter-analysis-1}.

\noindent \textbf{Lookback window size $T$.} 
We analyze the performance variation with different lookback window size $T$ in Figure \ref{figure-parameter-analysis-1} (a). Our model performs best when $T$ is set to about 20 in both datasets.

\noindent \textbf{Implicit relation threshold $\eta$.} 
The results of our model with different implicit relation thresholds are reported in Figure \ref{figure-parameter-analysis-1} (b). 
The performance of our proposed model grows with the increment of $\eta$ and achieves the best performance when $\eta$ is set to 0.0054 in CSI100E. With the increment of $\eta$, the performance raises at first and then drops gradually in the dataset CSI300E. When the $\eta$ becomes bigger, the performance decreases  possibly because some meaningful implicit edges are neglected.






\section{Conclusion and Future Work}
\label{section-conclusion}
In this paper, we focus on stock movement prediction task. To model stock momentum spillover in real financial market, we first construct a novel bi-typed hybrid market knowledge graph. Then, we propose a novel Dual Attention Networks, which are equipped with both inter-class attention module and intra-class attention module, to learn the stock momentum spillover features on the newly constructed MKG. To evaluate our method, we construct two new datasets CSI100E and CSI300E. The empirical experiments on the constructed datasets demonstrate our method can successfully improve stock prediction with bi-typed hybrid-relational MKG via the proposed \textsc{DanSmp}.
The ablation studies reaffirm that the performance gain mainly comes from the use of the associated executives, and additional implicit relation between companies in MKG.

An interesting future work direction is to explore web media about the executives including: (i) the negative facts from news, such as accusation of crime, health issue, etc; (ii) the improper speech on social media, such as Twitter and Weibo. We believe these factual event information of executives can be detected and utilized to feed into graph-based methods for better SMP performance.

\begin{acks}
The authors would like to thank all anonymous reviewers in advance.
This research has been partially supported by grants from the National Natural Science Foundation of China under Grant No. 71725001, 71910107002, 61906159, 62176014, U1836206, 71671141, 71873108, 62072379, the State key R \& D Program of China under Grant No. 2020YFC0832702, the major project of the National Social Science Foundation of China under Grant No. 19ZDA092, and the Financial Intelligence and Financial Engineering Key Laboratory of Sichuan Province.
\end{acks}

\bibliographystyle{ACM-Reference-Format}
\bibliography{sample-base}










\end{document}